\documentclass[floats,floatfix,amssymb,prd,twocolumn,superscriptaddress,nofootinbib,preprintnumbers]{revtex4-1}
		
\usepackage{amssymb,amsmath,verbatim,mathtools,needspace,enumitem,etoolbox,graphicx,physics,microtype,afterpage,bm}

\usepackage[utf8]{inputenc}
\usepackage{amsmath,amssymb}
\usepackage{amsfonts}
\usepackage{bm}
\usepackage{courier}
\usepackage{graphics,orcidlink}
\usepackage{float}
\usepackage{amsmath}
\usepackage{amssymb}
\usepackage[normalem]{ulem} 
\usepackage[mathscr]{euscript}
\usepackage{acronym}
\usepackage{float}
\usepackage[caption=false]{subfig}
\usepackage{lipsum}
\usepackage{mathtools}
\usepackage{multirow}
\usepackage{graphicx}
\usepackage{mathtools}
\usepackage{multirow}
\usepackage{graphicx}

\usepackage{ragged2e}
\DeclareCaptionJustification{justified}{\justifying}
\makeatletter
\newcommand{\subsetsim}{\mathrel{\mathpalette\subset@sim\relax}}
\newcommand{\subset@sim}[2]{%
  \vtop{\offinterlineskip\m@th
    \ialign{\hfil##\cr
      $#1\subset$\cr\noalign{\kern0.5pt}\scalebox{0.9}{$#1\sim$}\cr
    }%
  }%
}
\makeatother

\usepackage{amssymb,amsmath,verbatim,mathtools,needspace,enumitem,etoolbox,graphicx,physics,microtype,afterpage,bm}

\usepackage{multirow}
\usepackage{pifont}
\usepackage{lmodern}

\usepackage{multirow}

\allowdisplaybreaks
\usepackage{tikz}
\usepackage{color}
\usepackage{framed}
\usepackage{hyperref}
\hypersetup{colorlinks, citecolor=bluscuro, linkcolor=black, urlcolor=bluscuro}
\definecolor{rossos}{cmyk}{0,1,1,0.55}
\definecolor{bluscuro}{rgb}{0.15, 0.2, .85}
\definecolor{bluchiaro}{cmyk}{1,.3,0.,0.1}
\definecolor{ForestGreen}{rgb}{0.13, 0.55, 0.13}
\definecolor{azure}{rgb}{0.0, 0.5, 1.0}
 
\usepackage{slashed}
\usepackage{hyperref}

\newcommand{\apjl}{"Astrophys. J. Lett."}

\def\bea{\begin{eqnarray}}
\def\eea{\end{eqnarray}}

\def\d{{\mathrm{d}}}

\newcommand{\bs}{\begin{subequations}}
\newcommand{\es}{\end{subequations}}

\newcommand{\be}{\begin{equation}}
\newcommand{\ee}{\end{equation}}
\renewcommand{\d}{{\rm d}}

\def\lsim{\mathrel{\rlap{\lower4pt\hbox{\hskip0.5pt$\sim$}}
    \raise1pt\hbox{$<$}}}         
\def\gsim{\mathrel{\rlap{\lower4pt\hbox{\hskip0.5pt$\sim$}}
    \raise1pt\hbox{$>$}}}         

\newcommand{\unmezzo}{\frac{1}{2}}

\renewcommand{\d}{{\rm d}}

\makeatletter
\def\l@subsubsection#1#2{}
\makeatother

\newcommand{\sapienza}{Dipartimento di Fisica, Sapienza Università 
	di Roma, Piazzale Aldo Moro 5, 00185, Roma, Italy}
\newcommand{\infn}{INFN, Sezione di Roma, Piazzale Aldo Moro 2, 00185, Roma, Italy}
\newcommand{\nottingham}{Nottingham Centre of Gravity \& School of Mathematical Sciences, University of Nottingham, University Park, Nottingham, NG7 2RD, UK}

\begin{document}
\title{
Extreme mass-ratio inspirals as probes of fundamental dipoles
}

\begin{abstract}
Even if globally neutral, in various scenarios compact objects can have a nonvanishing dipole moment. Examples include neutron stars with magnetic dipoles, black-hole microstates in the string-theory fuzzball scenario, and classical black holes in modified theories of gravity with spin-induced scalarization or Lorentz-violating terms.
A fundamental dipole moment would give rise to rich phenomenology, for example to intrinsic precession and extra emission channels in binary systems.
We show that extreme mass-ratio inspirals~(EMRIs) detectable by future gravitational-wave interferometers allow us to study a fundamental dipole on the secondary object in a model-agnostic fashion.
By developing a general model for a fundamental scalar dipole, we compute the extra flux associated with it. This effect is suppressed by the square of the mass ratio relative to the case of fundamental charges, making its detection with EMRIs very challenging for the typical dipole moments predicted in various models. On the other hand, for the same reason the impact of an extra dipole for constraints on extra fundamental charges is likely negligible, making the latter constraints more robust.
\end{abstract}

\author{Jacopo Lestingi}
 \email{jacopo.lestingi@nottingham.ac.uk}
\affiliation{\sapienza}
\affiliation{\infn}
\affiliation{\nottingham}

\author{Enrico Cannizzaro\orcidlink{0000-0002-9109-0675}}
\email{enrico.cannizzaro@uniroma1.it}
\affiliation{\sapienza}
\affiliation{\infn}

\author{Paolo Pani\orcidlink{0000-0003-4443-1761}}
\email{paolo.pani@uniroma1.it}
\affiliation{\sapienza}
\affiliation{\infn}

\date{\today}
\maketitle

{
  \hypersetup{linkcolor=black}

\section{Introduction}\label{sec:intro}

The famous no-hair theorems predict that, in a large class of theories, black holes~(BHs) are described by the Kerr-Newman solution and do not have any extra charge other than the electromagnetic one.
Circumventing these no-go theorems has motivated both theoretical work --~aimed at finding theories in which BHs can have hair~-- and phenomenological work --~aimed at finding the consequences of this extra hair (see, e.g.,~\cite{Berti:2015itd,Barack:2018yly}). 
The most natural and best studied case is when BHs are endowed with extra fundamental charges, which give rise to dipolar radiation in binary systems and can be probed with binary pulsar timing and gravitational-wave~(GW) inspirals (see~\cite{Berti:2015itd} for a review).

The absence of dipolar emission in binary pulsars~\cite{Kramer:2021jcw} and in GW events~\cite{Barausse:2016eii,LIGOScientific:2021sio} already puts stringent constraints on the existence of fundamental charges in various contexts.

In the future, extreme mass-ratio inspirals~(EMRIs) --~ one of the main targets of future space detectors such as LISA~\cite{LISA:2017pwj}~-- will provide a unique probe to search for extra fundamental charges~\cite{Barausse:2020rsu,LISA:2022kgy}, either in the context of specific modified theories of gravity~\cite{Cardoso:2011xi,Yunes:2011aa,Pani:2011xj} or in a model-agnostic fashion, as recently shown~\cite{Maselli:2020zgv,Maselli:2021men,Barsanti:2022ana,Barsanti:2022vvl} for scalar fields (see Refs.~\cite{Liang:2022gdk,Zhang:2023vok} for extensions to the vector case).

In addition to new fundamental charges, there is strong theoretical and phenomenological motivation for models in which compact objects are \emph{globally neutral} (hence evading standard dipole-emission constraints) but can nevertheless have higher multipole moments.
The most natural example are magnetars, which are endowed with strong magnetic dipole moments (see~\cite{Kaspi:2017fwg} for a review).
Furthermore, in the context of modified gravity theories, BHs could have a fundamental dipole moment in Lorentz-violating theories~\cite{Barausse:2011pu}, in dynamical Chern-Simons gravity~\cite{Alexander:2009tp,R:2022tqa}, and in theories featuring spin-induced spontaneous scalarization~\cite{Dima:2020yac,Berti:2020kgk,Herdeiro:2020wei,Elley:2022ept} (see~\cite{Doneva:2022ewd} for a recent review).
Finally, in the context of BH microstates emerging in the string-theory fuzzball scenario~\cite{Mayerson:2020tpn,Bena:2022rna,Bena:2022ldq}, a long-lasting problem is to find consistent solutions which are globally neutral. Remarkably, this was recently achieved with topological solitons~\cite{Bah:2020ogh,Bah:2021owp}, which are globally neutral but have an intrinsic dipole moment.

Motivated by the above scenarios in various contexts, in this paper we wish to study the impact of fundamental dipoles for GW tests of fundamental physics with EMRIs.
Henceforth we use $G=c=1$ units, except for plots, where we restore the speed of light $c$ for clarity.

\section{Setup}\label{sec:setup}

\subsection{Theoretical framework}
Let us consider the following generic action~\cite{Maselli:2020zgv}
\be
 S[\mathbf{g},\Phi, \psi] = S_0[\mathbf{g}, \Phi] +\alpha S_c[\mathbf{g},\Phi] + S_m[\mathbf{g}, \Phi, \psi],
\ee
where $\Phi$ is a massless scalar field, 
\be
 S_0[\mathbf{g}, \Phi] = \int \d^4 x \frac{\sqrt{-\mathbf{g}}}{16 \pi} (R - \unmezzo \partial_{\mu} \Phi\partial^{\mu}\Phi),
\ee
$S_m$ is the action of the matter fields $\psi$, while the action $S_c$ describes a generic non-minimal coupling between gravity and the scalar field, whose coupling constant is $\alpha$. As in Refs.~\cite{Maselli:2020zgv, Maselli:2021men, Barsanti:2022ana, Barsanti:2022vvl},  
we will assume that the theory is continuously connected to GR in the $\alpha \to 0$ limit and that either $\alpha$ has dimensions $[\alpha]=({\rm mass})^n$ with $n\geq 1$ or that the theory is such that no-hair theorems hold.

An EMRI is a binary system in which a small compact object with mass $m_{SCO} = 2 \mu$ is spiraling around a supermassive BH with mass $M\gg\mu$. Owing to the small mass ratio, $q=\mu/M\ll1$, one can model the secondary using the "skeletonized approach"~\cite{1975ApJ...196L..59E, Damour:1992we, Julie:2017ucp}, in which the secondary object is treated as a point particle. Nevertheless, as we want to describe an object endowed with a dipolar field, we will use the skeletonized approach to model the secondary as an elementary dipole made by two point particles with mass $\mu$ displaced by a constant separation $\delta y^{\mu}$, which we assume to be small with respect to the length scale of the exterior spacetime, $\sim M$. 
Therefore, the action of the matter fields reduces to
\begin{equation}
      S_m[\mathbf{g}, \Phi]= - \sum_{i=1,2} \int \d \lambda m_i(\Phi) \sqrt{\mathbf{g}_{\mu \nu} \frac{\d y_i^{\mu}}{\d \lambda} \frac{\d y_i^{\nu}}{\d \lambda} },
\end{equation}
where the two worldlines of the particles are given by $y_{1,2}^{\mu}(\lambda) = y^{\mu}(\lambda) \pm \unmezzo \delta y^{\mu}$ with $y^{\mu}(\lambda)$ world-line of the center of mass of the dipole. 

From the action, we can now derive the field equations and solve them by perturbatively expanding the fields at the leading order in the mass ratio. The Einstein equations for the gravitational field read
\begin{equation}
\label{eq:EinsteinEq}
    G_{\mu \nu} = R_{\mu\nu}-\frac{1}{2}\mathbf{g}_{\mu\nu}R= T_{\mu \nu}^{(s)}+\alpha T_{\mu \nu}^{(c)}+T_{\mu \nu}^{p},
\end{equation}
where $T_{\mu \nu}^{(s)}$ is the stress energy tensor of the scalar field,  $T_{\mu \nu}^{(c)}$ is the term arising from the variation of the non-minimal coupling term $S_c$, while $T_{\mu \nu}^{p}$ is the stress-energy tensor of the two-particle dipole,
\begin{equation}
\label {eq:Tmunu}
      T^{p \ \mu \nu}= 8\pi \sum_{i=1,2} \int \d \lambda m_i(\Phi) \frac{\delta^{(4)}(x^{\alpha}- y_i^{\alpha}(\lambda))}{\sqrt{-\mathbf{g}}} \frac{\d y_i^{\mu}}{\d \lambda} \frac{\d y_i^{\nu}}{\d \lambda}.
\end{equation}
Varying the action with respect to the scalar field yields
\be
\label{eq:ScalarField}
\Box \Phi + \frac{8 \pi \alpha}{\sqrt{-\mathbf{g}}} \frac{\delta S_c}{\delta \Phi} = 16\pi \sum_{i=1,2} \int \d \lambda  m'_i(\Phi) \frac{\delta^{(4)}(x^{\mu}- y_i^{\mu}(\lambda))}{\sqrt{-\mathbf{g}}}\,,
\ee
where the prime denotes derivative of a function with respect to its argument.

As discussed in~\cite{Maselli:2020zgv,Maselli:2021men}, due to the mass dimensions of the coupling $\alpha$, GR modifications to the background are suppressed by the mass ratio of the binary (or absent if the no-hair theorems are satisfied).
Hence, the exterior spacetime of the primary can be approximated as (or is exactly) the Kerr metric.
Furthermore, in these settings, one can neglect the terms proportional to $\alpha$ in Eqs.~\eqref{eq:EinsteinEq} and~\eqref{eq:ScalarField}, since they are suppressed by the mass ratio.
In absence of the secondary, the resulting set of equations coincide with those of general relativity with a free scalar field, for which the no-hair theorem applies. Therefore, the background scalar field is just a constant $\Phi_0$.

By expanding~\eqref{eq:ScalarField} at linear order $\Phi= \Phi_0 + \varphi$, we obtain the following equation for the perturbation $\varphi$:
 \begin{equation}
 \label{eq:phi}
     \Box \varphi= 16\pi \sum_{i=1,2} \int \d \lambda  m'_i(\Phi_0) \frac{\delta^{(4)}(x^{\mu}- y_i^{\mu}(\lambda))}{\sqrt{-\mathbf{g}}} \, ,
\end{equation}
where the operator $\Box$ is evaluated on the background Kerr metric $\mathbf{g}^0_{\mu \nu}$. The same expansion leads $m_i(\Phi)$ in Eq.~\eqref{eq:Tmunu} to be evaluated at $\Phi = \Phi_0$.  Thus, at the leading order the gravitational equations ~\eqref{eq:EinsteinEq} coincide with the standard ones for two infinitely close point masses in general relativity.

Let us now discuss the physical meaning of the terms $m_i(\Phi_0)$ and $m'_i(\Phi_0)$ by generalizing the argument of Ref.~\cite{Maselli:2021men} to the case of our two-particle system. These functions can be evaluated in a region which is sufficiently close to the particles (relatively to the length scale of the exterior spacetime, $\sim M$), but sufficiently far away from them in the length scale of the particle themselves, $\sim \mu$, so that we can evaluate the equations in the weak-field limit. 
We therefore choose a reference frame $\{\Tilde{x}^{\mu}\}$ centered at the center of mass of the compact object and consider Eqs.~\eqref{eq:Tmunu} and~\eqref{eq:phi} in an intermediate region, $\mu \ll \Tilde{r} \ll M$, where $\Tilde{r}^2 = \Tilde{x}^i \Tilde{x}_i  $. Let us first consider Eq.~\eqref{eq:Tmunu} evaluated at $\Phi = \Phi_0$. As in this region we are in the weak-field limit, the stress-energy tensor of a particle reduces to its matter density and thus it follows that $m_1(\Phi_0)=m_2(\Phi_0)=\mu$. 

We can now turn to the study of Eq.~\eqref{eq:phi}. Expanding the latter to leading order in the infinitesimal displacement $\delta y^\mu\ll x^\mu$ in these coordinates yields
 \begin{align}
       \nabla^2 \varphi & = 16\pi A(\Phi_0) \delta^{(3)}(\Tilde{x}^i) \nonumber \\ &
       +  16\pi B(\Phi_0) \delta \Tilde{y}^i \partial_i  \delta^{(3)}(\Tilde{x}^j),
 \end{align}
where $A(\Phi_0)= \unmezzo (m'_1(\Phi_0) + m'_2(\Phi_0)) $ and $B(\Phi_0)= \unmezzo (m'_1(\Phi_0) - m'_2(\Phi_0)) $. If $B(\Phi_0)=0$, then the solution $\varphi$ has exactly the same form as in~\cite{Maselli:2021men}, and therefore $A(\Phi_0)$ can be associated with the monopolar scalar charge per unit mass of the object (we shall denote this quantity by $d$). As we instead wish to describe an intrinsically dipolar field configuration, for the moment we ignore this term.
Setting $A(\Phi_0)=0$ we recognize the equation for the potential of a dipole, which is solved by
\be
\varphi = \frac{4B(\Phi_0) \delta \Tilde{y}^i \Tilde{x}_i}{\Tilde{r}^3}.
\ee
By direct comparison with the potential of a dipole with dipole vector $P^i = \mu d \delta \Tilde{y}^i$, which is
\be
\varphi = \frac{ P^i \Tilde{x}_i}{\Tilde{r}^3},
\ee
it is clear that we can interpret $4B(\Phi_0) \delta \Tilde{y}^i $ as a dipole vector and therefore $B(\Phi_0)= \frac{1}{4} \mu d$, $m'_1(\Phi_0)=-m'_2(\Phi_0)= \frac{1}{4} \mu d$ where, as mentioned, $d$ is the scalar charge per unit mass of the secondary.
Finally, the equation for the scalar field reads
\be
\Box \varphi = 4\pi T \, , \label{eq:finally}
\ee
where
\begin{align}
    \label{eq:T}
T & = \mu d \left [ \int \d \lambda \frac{\delta^{(4)}(x^{\mu}- y^{\mu}(\lambda)+ \frac{\delta y^{\mu}}{2})}{\sqrt{-\mathbf{g}}} \right . \nonumber \\ & - \left . \int \d \lambda \frac{\delta^{(4)}(x^{\mu}- y^{\mu}(\lambda)- \frac{\delta y^{\mu}}{2})}{\sqrt{-\mathbf{g}}} \right ].
\end{align}

The above discussion shows that our system can indeed be understood as a scalar dipole made of two particles with the same mass but opposite scalar charge, whose center of mass inspirals onto a standard supermassive (Kerr) BH. This suggests to introduce the dipole moment tri-vector per unit mass squared:  
\be
\label{eq:DipoleMoment}
p^{i} = \frac{d}{\mu} \delta y^{i}.  
\ee 
Notice that this is a dimensionless quantity, in analogy with the dimensionless charge $d$.  

In the case in which $A(\Phi_0) \neq 0$, then the secondary has a nonvanishing net charge as well as a dipole moment. 
While we are mostly interested in the case of zero net charge, later on we will also consider this scenario, as our formalism allows analysing deviations from scalar emission from a fundamental charge due to the presence of an extra dipole component. 

\subsection{Scalar equation via Teukolsky formalism}
The inhomogeneous Klein-Gordon equation~\eqref{eq:finally} can be solved via Teukolsky formalism~\cite{Teukolsky:1973ha}. First of all, we must characterize the wordline $y^\mu(\lambda)$ of the center of mass, that appears on the right-hand side of Eq.~\eqref{eq:finally}. Since for an EMRI the inspiral timescale is much longer than the orbital time scale, $T_{\rm inspiral} \gg T_{\rm orbital}$, we can adopt an adiabatic approximation, which allows us to consider the center of mass of the dipole as being in nearly geodesic motion.
This approximation facilitates the evaluation of the emitted energy flux $\dot{E}$ from the inspiral at each time. For simplicity, we will consider equatorial, circular orbits of the Kerr metric and use Boyer-Lindquist coordinates
$\{t,r,\theta,\phi\}$. The geodesic of the centre of mass in this setting is described by the following constants of motion, which describe the energy, angular momentum, and angular velocity of the center of mass, respectively,
\begin{eqnarray}
    E_c &=& \frac{a \sqrt{M} + \sqrt{r_0}(r_0 - 2M)}{r_0^{3/4} \sqrt{2a \sqrt{M} + \sqrt{r_0}(r_0 - 3M)}},\\
    L_c &=& \frac{\sqrt{M}(r_0^2-2a\sqrt{Mr_0} + a^2)}{r_0^{3/4} \sqrt{2a \sqrt{M} + \sqrt{r_0}(r_0 - 3M)}},\\
    \Omega_c &=& \frac{\sqrt{M}}{a\sqrt{M} + r_0^{3/2}}\,,
\end{eqnarray}
where $r_0$ is the orbital radius of the geodesic and $Ma$ is the angular momentum of the Kerr BH. Hence, the wordline of the center of mass is $y^\mu(\lambda)=(t_p(\lambda), r_0, \pi/2, \Omega_c t_p(\lambda))$. 

We will assume that the displacement $\delta y^\mu$ is constant, i.e. it does not depend on the affine parameter $\lambda$. Note that in general, if the displacement is not aligned with the spin of the primary, the interaction between the latter and the scalar dipole will induce precession even in the case of initially circular and equatorial orbits. Hence, for generic orientations of the dipole moment, our assumption of a constant displacement is only valid on timescales that are much shorter than the precession timescale, such that this effect can be neglected. The precession timescale is shorter than, or at most comparable to for relativistic orbits, the inspiral timescale, but much larger than the orbital one, i.e. $T_{\rm inspiral} \gg T_{\rm precession} \gg T_{\rm orbital}$, see e.g.~\cite{Stavridis:2009mb}. Hence, while our formalism does not allow us to consistently evolve the binary through the entire inspiral (for generic displacements), it safely allows us to evaluate the scalar fluxes throughout the orbital motion.
Of course this limitation is absent if the displacement $\delta y^\mu$ is orthogonal to the equatorial plane, since there is not precession in that case.

With this in mind, we can simplify the field equation~\eqref{eq:finally} by expanding the trace of the stress energy tensor with respect to the constant displacement between the two particles, $\delta y^\mu=(\delta t, \delta r, \delta \theta, \delta \phi)$. This yields
\be
\label{eq:Texplicit}
T \simeq \frac{1}{\sqrt{-\mathbf{g}}}(\delta t \partial_t + \delta r \partial_r + \delta \theta \partial_{\theta} + \delta \phi \partial_{\phi})(\sqrt{-\mathbf{g}}T_p)\,,
\ee
where $T_p$ has the same expression of the source of the scalar field in the setting in which the secondary is endowed with a scalar monopolar charge:
\be
T_p = \frac{\mu d}{\Sigma \sin{\theta} \abs{\dot{t_p}}} \delta(r-r_0) \delta(\theta - \frac{\pi}{2}) \delta (\phi -\Omega_c t) \,,
\ee
with $\Sigma = r^2 + a^2 \cos^2\theta$ and $\dot{t}_p=d t_p(\lambda)/d\lambda $.
Finally, we can perform a Fourier transform and expand both the scalar field and the source in spin-weighted spheroidal harmonics
\begin{align}
\varphi (t,r,\Omega) &= \int \d \omega \sum_{l,m} \frac{X_{lm}(r,\omega)}{\sqrt{r^2+a^2}} S_{0lm}(\theta, \omega) e^{im\phi} e^{-i\omega t}\,,\\
4\pi \Sigma T &= \int \d \omega \sum_{l,m} T_{lm}(r,\omega) S_{0lm}(\theta, \omega) e^{im\phi} e^{-i\omega t}\,.\label{eq:expansionT}
\end{align}
This decomposition allows us to decouple the angular and radial dependence of the scalar field. Indeed, we obtain the standard inhomogeneous differential equation for the radial field $X_{lm}(r,\omega)$:
\be
\label{eq:non-homogeneous}
\left[ \frac{\d^2}{\d r_*^2} + V \right] X_{lm}(r,\omega) = \frac{\Delta}{(r^2 + a^2)^{3/2}} T_{lm}(r,\omega)\,,
\ee
where $V$ is the effective potential and can be found, for example, in~\cite{10.1143/PTP.96.713}.
We can obtain $T_{lm}$ as a function of $T$ by inverting Eq.~\eqref{eq:expansionT} using the properties of the spheroidal harmonics:
\be
\label{eq:genericTlm}
T_{lm}(r, \omega) = 2\int \d t \d \theta \d \phi \Sigma \sin{\theta} T S^*_{0lm} e^{-im\phi} e^{i\omega t}\,.
\ee
We can now substitute Eq.~\eqref{eq:Texplicit} into Eq.~\eqref{eq:genericTlm} and perform the integrals in $\theta$ and $\phi$, through integration by parts and the properties of the $\delta$ function. We finally obtain
\begin{align}
     T_{lm} & = \frac{4\pi \mu d}{\abs{\dot{t}}} \delta (m\Omega_c - \omega) \left[  
  S^*_{0lm}(\frac{\pi}{2}, m \Omega_c ) \delta r \partial_r \delta(r-r_0) \right. \nonumber \\ &
  - \frac{\d}{\d \theta} S^*_{0lm}(\frac{\pi}{2},m \Omega_c) \delta \theta \delta (r-r_0) \nonumber \\ & + \left. (\delta \phi - \Omega_c \delta t) S^*_{0lm}(\frac{\pi}{2},m \Omega_c) i m \delta (r-r_0)  \right]. \label{eq:Tlm}
 \end{align}
%
%
Note that the displacements along the $t$ and $\phi$ directions are proportional to each other\footnote{
Note also that both terms are proportional to the azimuthal number $m$. While this is obvious for the derivative with respect to $\phi$, 
it arises also for the time derivative because the latter brings a factor $\omega$, and circular motion implies $\omega=m\Omega_c$.}.
This is a consequence of the chosen equatorial circular motion. In fact, for circular equatorial orbits, performing a displacement $ \delta \phi$ corresponds to moving the particle along the orbit by an angle proportional to $\Omega_c \delta t$ and vice-versa. In particular, for \textit{prograde} orbits such that $\Omega_c>0$, by considering a positive displacement $\delta \phi= \Omega_c \delta t$ the sum of these two terms is zero. This is because the angular term $\delta \phi$, if positive, shifts the particle along the circular orbit in a clockwise way. A shift $\delta t$ along time instead, corresponds to $\phi= \Omega_c(t+ \delta t)$, i.e. $\phi-\Omega_c\delta t = \Omega_c t$, so that at time $t$ the particle is actually displaced by an angle $-\Omega_c\delta t= - \delta \phi$ with respect to its original position in the counterclock direction. For \textit{retrograde} orbits, the effect is clearly reversed. In general, as these terms are proportional, we can neglect from now on shifts along time, and simply re-absorb them as shifts along the $\phi$ direction.

\section{Dipole-induced scalar fluxes}
\subsection{Analytic derivation}
The solution of the inhomogeneous equation~\eqref{eq:non-homogeneous} can be found using the standard Green function in terms of two independent solutions of the corresponding homogeneous equation.
The latter have the following asymptotic behavior:
 \begin{equation}
 \label{eq:solutionhomogeneous}
     X_{lm\omega}^{\infty, r_+}\sim e^{\pm i r_* k_{\infty, +}} \ \text{as} \ r \rightarrow \infty, r_+
 \end{equation}
where $k_+=\omega - m \Omega_H$, $\Omega_H$ being the angular velocity of the BH horizon, and $k_\infty=\omega$. The solution of the inhomogeneous equation reads
        \begin{align}
        \label{eq:solutioninhomogeneous}
             X_{lm \omega}(r) & = W^{-1} X^{\infty}_{lm \omega} \int_{r_+}^{r} \mathrm{d} s \frac{T_{lm \omega} X^{r_+}_{lm \omega}}{\sqrt{s^2 + a^2}} \nonumber \\ &  +  W^{-1} X^{r_+}_{lm \omega} \int_{r}^{\infty} \mathrm{d} s \frac{T_{lm \omega} X^{\infty}_{lm \omega}}{\sqrt{s^2 + a^2}}
            \,,
        \end{align}
where $W$ is the Wronskian of the two homogeneous solutions. 
To evaluate the fluxes we are interested in the asymptotic behaviour of the solution~\eqref{eq:solutioninhomogeneous} at infinity and at the horizon. Using Eqs.~\eqref{eq:Tlm} and~\eqref{eq:solutionhomogeneous}, we get
        \begin{align}
        X^{\rm out}_{lm \omega} &= Z^{\infty}_{lm \omega} \delta (m\Omega_c - \omega) e^{i m\Omega_c r_{*} }, \,,\\
        X^{\rm in}_{lm \omega} &= Z^{r_+}_{lm \omega} \delta (m\Omega_c - \omega) e^{i m(\Omega_c - \Omega_H) r_{*} }. 
        \end{align}
where  
         \begin{equation}
         \label{eq:asymptoticbehaviour}
             Z^{\infty, r_+}_{lm \omega}  = \left. W^{-1} \frac{4\pi \mu d}{\abs{\dot{t}}} (\mathcal{R} + \Theta + \Tilde{\Phi}) \frac{X^{r_+,\infty}_{lm \omega}(r)}{\sqrt{r^2 + a^2}}\right \rvert_{r=r_0},
        \end{equation}
with
         \begin{align}
         \label{eq:radialterm}
   \mathcal{R} &= S^*_{0lm}(\frac{\pi}{2}, m \Omega_c ) \delta r \partial_r , \\
  \label{eq:thetaterm}
      \Theta &=  - \frac{\mathrm{d}S^*_{0lm}}{\mathrm{d} \theta} (\frac{\pi}{2},m \Omega_c) \delta \theta ,\\
    \label{eq:phiterm}
      \Tilde{\Phi} &=  i m S^*_{0lm}(\frac{\pi}{2},m \Omega_c) \delta \phi .
  \end{align}

The fluxes at infinity and at horizon can be computed from the $t-r$ component of the scalar field stress-energy tensor,
        \begin{equation}
          F_{\infty, r_+} = \frac{\mathrm{d}E_{\infty,r_+}}{\mathrm{d}t} = \lim_{r \to \infty , r_+}
            \int \mathrm{d}\Omega \left. \Sigma T^{(s)} \right. ^r_{t}\,.
        \end{equation}

Finally, using Eq.~\eqref{eq:asymptoticbehaviour}, the fluxes read:
\begin{align}
\label{eq:flux_infty}
     F_{\infty} &= \frac{\mathrm{d}E_{\infty}}{\mathrm{d}t} = \sum_{l,m} \abs{Z^{\infty}_{lm \omega}}^2 m^2 \Omega_c^2 , \\
\label{eq:flux_hor}
      F_{r_+} &= \frac{\mathrm{d}E_{r_+}}{\mathrm{d}t} = \sum_{l,m} \abs{Z^{r_+}_{lm \omega}}^2 m^2 \Omega_c (\Omega_c - \Omega_H). 
\end{align}
In the next section, we will compute numerically the dipole-induced scalar fluxes and highlight the differences with respect to the monopolar case. 
For an immediate comparison, we report here also the quantities $\Tilde{Z}^{\infty, r_+}_{lm \omega}$ characterizing the monopolar case:
   \begin{equation}
   \label{eq:Zmonopolar}
             \Tilde{Z}^{\infty, r_+}_{lm \omega}  = \left. W^{-1} \frac{4\pi \mu d}{\abs{\dot{t}}} 
              S^*_{0lm}(\frac{\pi}{2}, m \Omega_c ) \frac{X^{r_+,\infty}_{lm \omega}(r)}{\sqrt{r^2 + a^2}}\right \rvert_{r=r_0}.
        \end{equation}  

Even by a first qualitative analysis, a crucial difference can already be highlighted from dimensional considerations. From Eq.~\eqref{eq:asymptoticbehaviour}, one sees that the inhomogeneous solution, both at the horizon and at infinity, depends on the sum of three terms proportional to the components of the dipole moment vector, $ \delta r d/ \mu,  \delta\theta d/ \mu,  \delta\phi d/ \mu$, respectively.
First, we point out that the radial displacement is of the order of the secondary length scale, $\delta r = {\cal O}(\mu)$. Moreover, the term~\eqref{eq:radialterm} features $\partial_r$ which is ${\cal O}(M^{-1})$. As for the angular displacements, they can be roughly estimated as $\delta \theta, \delta \phi \sim \text{arcsin} (\mu/r_0)\approx \mu/r_0 \lesssim \mu/M$, from which $\delta \theta, \delta \phi = {\cal O}(q)$.  Thus, the terms~\eqref{eq:radialterm},~\eqref{eq:thetaterm},~\eqref{eq:phiterm} are all of the order ${\cal O}(q)$.
This immediately tells us that Eq.~\eqref{eq:asymptoticbehaviour} scales as the mass ratio squared, $q^2$, and the fluxes are proportional to $q^4$. 

In the case of a monopole instead, the fluxes are proportional to $q^2$. Therefore, \emph{emission from a dipole is intrinsically suppressed by a factor of $q^2$}. 

If we instead consider a configuration where both a scalar monopole and a scalar dipole are present (as we shall do in Sec.~\ref{sec:mixed}), the fluxes will be proportional to the square of the sum of the two terms~\eqref{eq:Zmonopolar} and~\eqref{eq:asymptoticbehaviour}. Hence, due to the double product of these two terms, the lowest order correction to the scalar monopolar flux due to the presence of the dipole will be proportional to $q^3$. This resembles the contribution of the spin of the secondary compact object to the gravitational fluxes~\cite{Piovano:2020zin,Piovano:2021iwv,Piovano:2022fsy}, which also enters at next-to leading order in the mass-ratio.

\subsection{Numerical results}
In this section we discuss the numerical results for the fluxes in the purely dipolar case. We consider an EMRI around a nearly-extremal Kerr BH with spin $a=0.99 M$. As we shall discuss, even if this choice maximizes the effect of the dipole, the latter is typically negligible. In all cases we compute the fluxes by summing the multipolar contributions up to $l=17$ and for all $m=-l,-l+1,...,l-1,l$.

The GW emission is studied by varying the radial coordinate $r_0$ of the center of mass of the secondary, and therefore its tangential velocity $v=\Omega_c r_0$, which increases as $r_0$ decreases. We will confront the dipole fluxes with the ones obtained in the monopolar case in the same setting~\cite{Yunes:2011aa, Maselli:2021men}. 

Recalling that, without loss of generality, we can set $\delta t = 0$, we can focus on the fluxes given by the three independent orientations for the dipole, which are $p^{i} = d (\delta r, 0, 0) / \mu$, $p^{i}= d (0, \delta \theta, 0) / \mu$, and $p^{i} = d (0, 0, \delta \phi) / \mu$, respectively (see Fig.~\ref{fig:draw}).


\begin{figure}[ht]
    \centering
    \includegraphics[width=9cm]{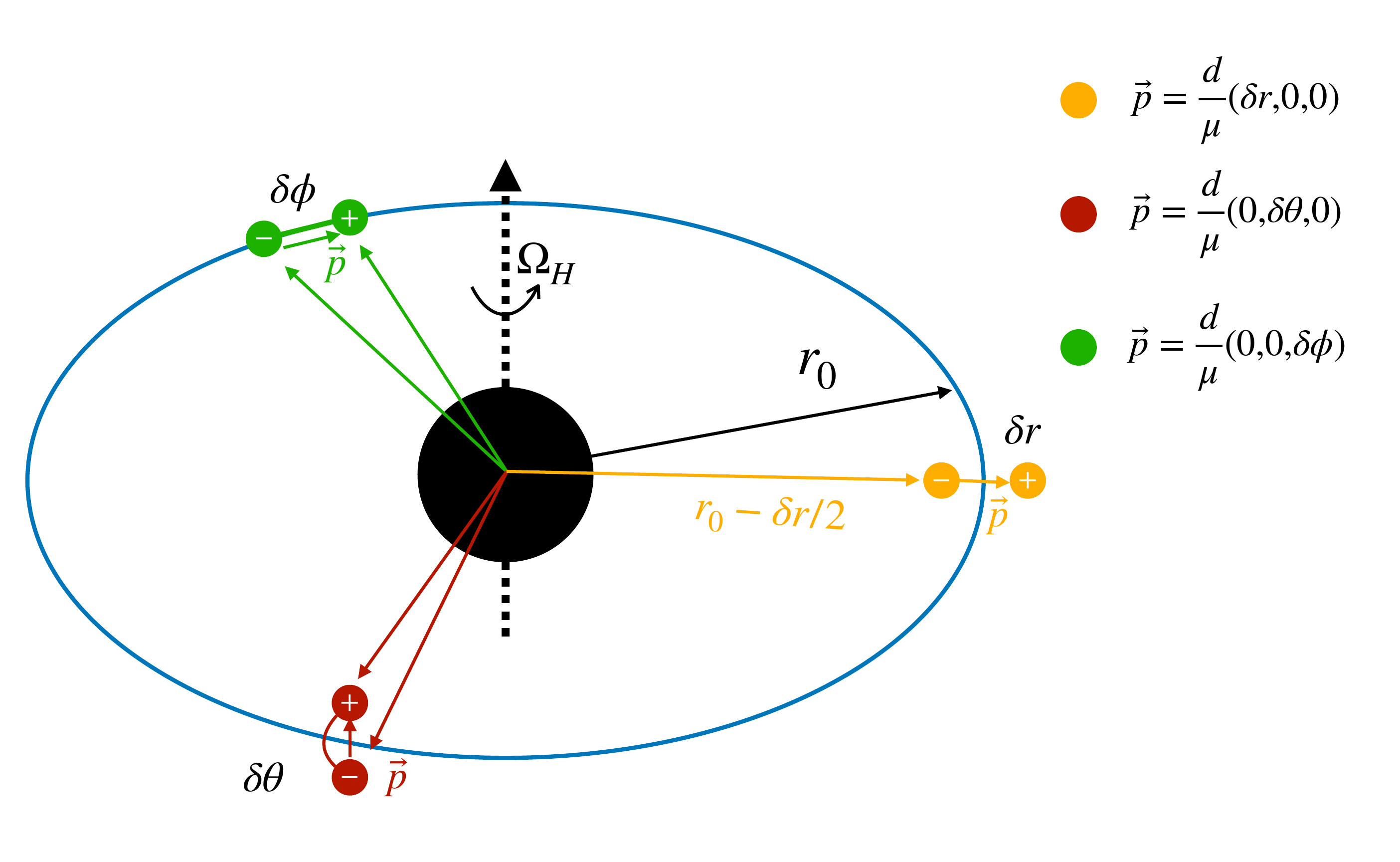}
    \caption{Schematic representation of an EMRI with a fundamental secondary dipole moment $\vec{p}$. We show the the three independent orientations of the dipole considered in this work. A generic orientation can be expressed as a linear combination of these three.}
    \label{fig:draw}
\end{figure}

Given that the mass ratio enters the fluxes as an overall factor, it is convenient to normalize both the dipolar and monopolar fluxes in a suitable way. 
For a given orientation of the dipole, we can define the normalized fluxes $\mathcal{F}_D$ and $\mathcal{F}_M$ as
  \begin{equation}
 \label{NormFluxes}
           F_M = (4\pi)^2 q^2 d^2  \mathcal{F}_M,
        \qquad
            F_D = (4\pi)^2 q^4 p^2  \mathcal{F}_D.
        \end{equation}
Here $F_D$ ($F_M$) is the flux obtained from Eqs.~\eqref{eq:flux_infty} and~\eqref{eq:flux_hor} using the $Z_{lm \omega}$ given in Eq.~\eqref{eq:asymptoticbehaviour} (Eq.~\eqref{eq:Zmonopolar}) for the dipolar (monopolar) case.
Also, $p$ is the magnitude of the dimensionless dipole vector, see Eq.~\eqref{eq:DipoleMoment}.
Using this normalization, $\mathcal{F}_D$ and $\mathcal{F}_M$ are independent of the dipole moment, charge, and mass ratio. However, for the purpose of a comparison one should keep in mind that, for the physical fluxes, ${F}_D$/${F}_M={\cal O}(q^2)\ll1$.

Figure~\ref{fig:Theta&Mono} shows the behaviour of normalized dipole emission (solid curves) in the $p^i = d  (0, \delta \theta, 0) / \mu$ case (in which the dipole is orthogonal to the equatorial plane and therefore precession in absent) in comparison to the normalized monopolar emission (dashed curves). In this configuration, for any orbit, the dipole emits a normalized flux that is always smaller than in the monopolar setup at least by one order of magnitude, both at horizon and at infinity. This is due to the dependence on $ \frac{\mathrm{d}S^*_{0lm}}{\mathrm{d} \theta} (\frac{\pi}{2},m \Omega_c)$ in the dipole emission. Indeed, it is easy to see that when the derivative of the spheroidal harmonic is computed at $\theta=\pi/2$, the (polar) $m=\pm l$ contribution to the fluxes, which usually is the dominant one, is identically zero. The first non-zero contribution comes from the axial mode $m=\pm(l-1)$, which is typically smaller.

\begin{figure}[ht]
    \centering
    \includegraphics[width=8cm]{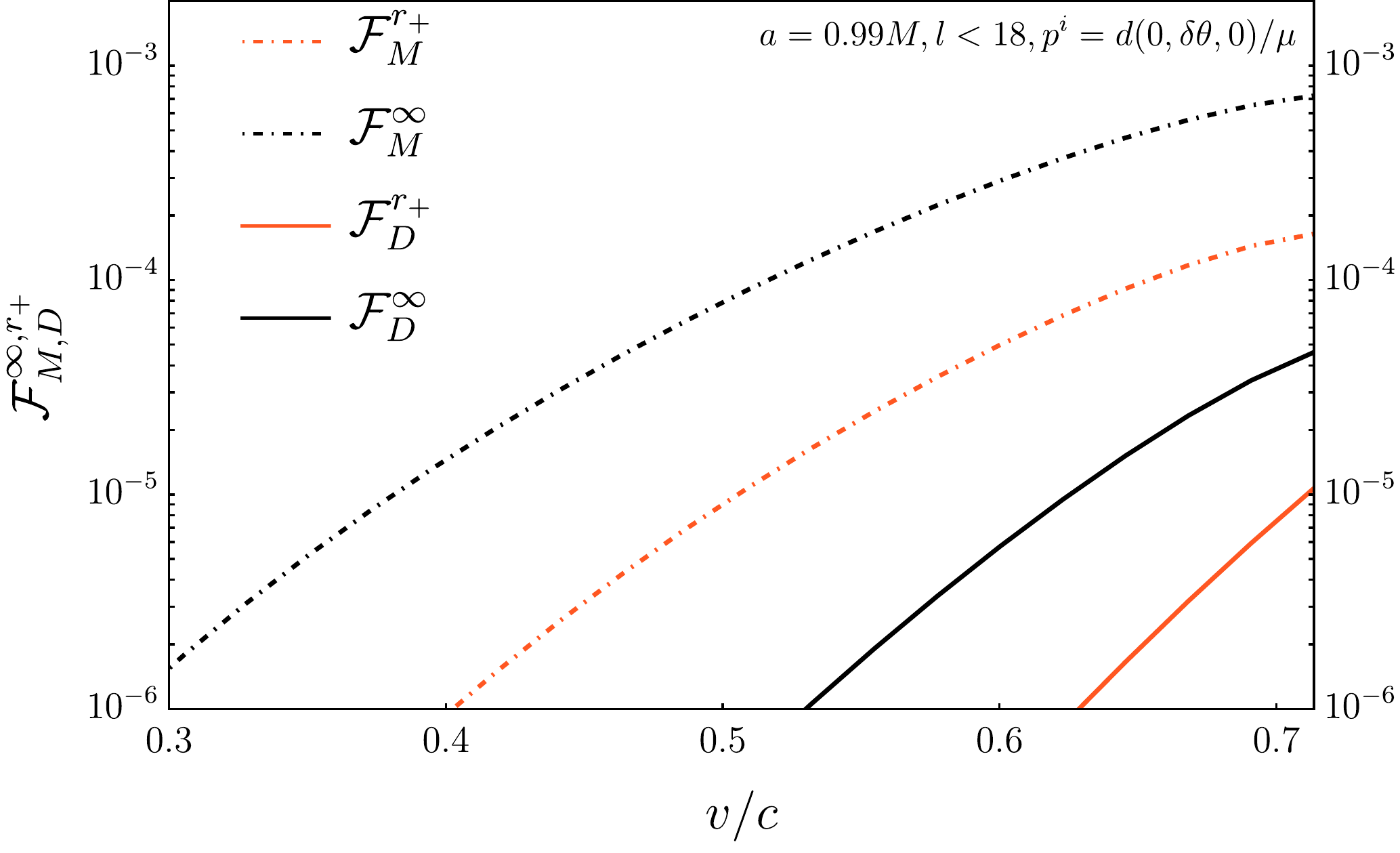}
    \caption{Solid curves: normalized scalar emission at infinity (black) and at horizon (orange) for a fundamental dipole with normalized moment $p^i = d (0, \delta \theta, 0)/ \mu$, namely the case in which the dipole is parallel to the spin of the primary BH. Dashed curves: for comparison we show the normalized scalar emission at infinity (black) and at horizon (orange) for the case of a fundamental charge.}
    \label{fig:Theta&Mono}
\end{figure}

In Fig.~\ref{fig:Radial&Mono} we show the behaviour of normalized dipole emission (solid curves) in the $p^i = d (\delta r, 0, 0) / \mu$ setting, where the dipole moment lays on the equatorial plane in radial direction.
We notice that the relative importance of the dipolar flux with respect to the monopolar one increases in regions of the spacetime with a stronger gravitational field. The normalized flux at infinity peaks at the innermost-stable circular orbit~(ISCO), where it is larger than the normalized monopolar one by almost two orders of magnitude. 

Figure~\ref{fig:Phi&Mono} shows the normalized dipole emission (solid curves) in the $p^i =  d (0, 0, \delta \phi) / \mu$ setup. This trend shown in this plot is similar to the previous one.
Overall the dipolar flux in this setting is always smaller than in the case of radially displaced dipole, but significantly larger than in the case of dipole aligned with the BH spin (Fig.~\ref{fig:Theta&Mono}).

Note that, as expected, in all three cases the fluxes increase as the small compact object gets closer to the ISCO since relativistic effects are amplified.

In the next section we will discuss the possible detectability of these fluxes, after restoring the normalization factors in Eq.~\eqref{NormFluxes}.

\begin{figure}[ht]
    \centering
    \includegraphics[width=8cm]{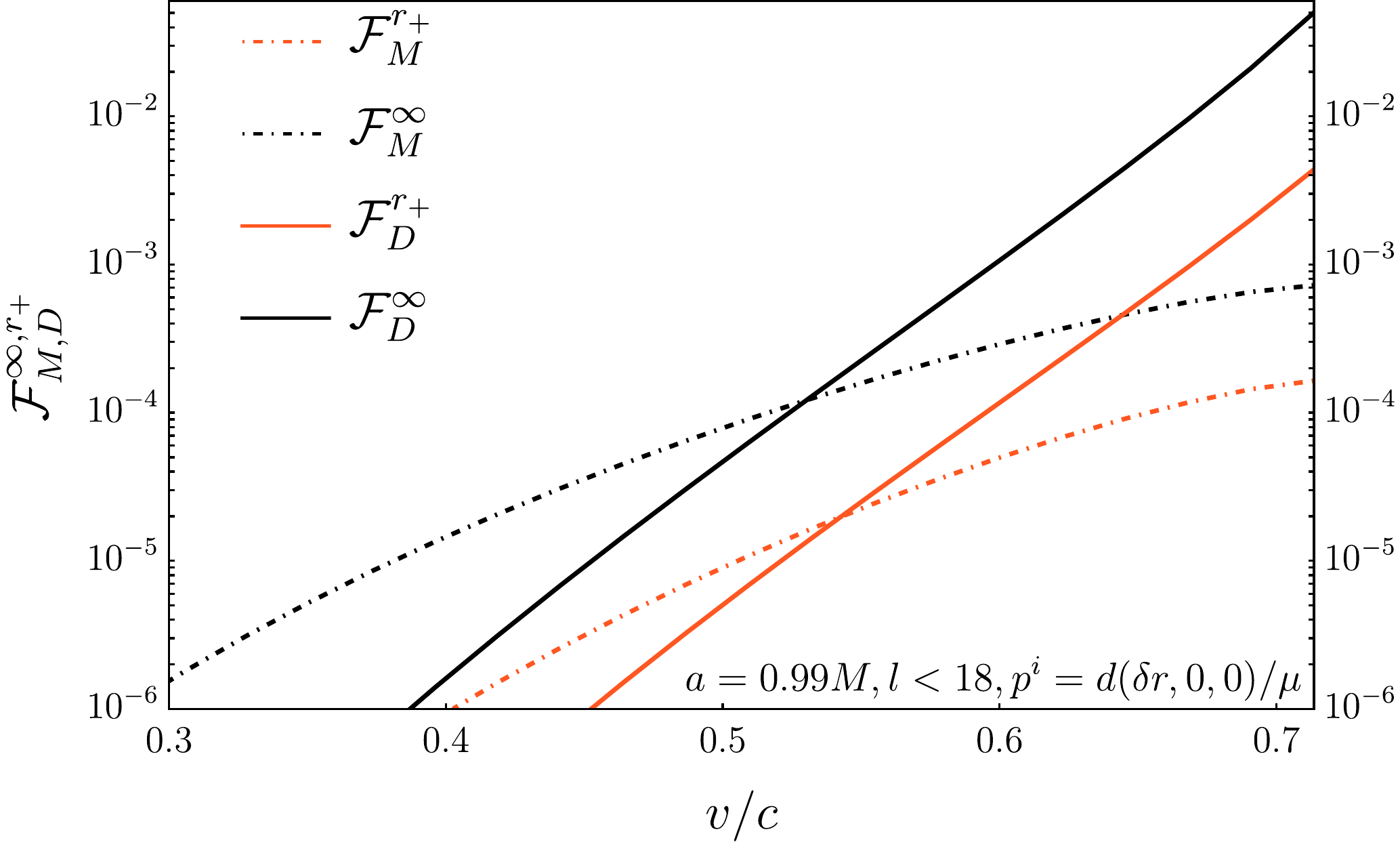}
    \caption{Same as Fig.~\ref{fig:Theta&Mono} but for for the case of a radial dipole moment along the equatorial plane.}
    \label{fig:Radial&Mono}
\end{figure}

\begin{figure}[ht]
    \centering
    \includegraphics[width=8cm]{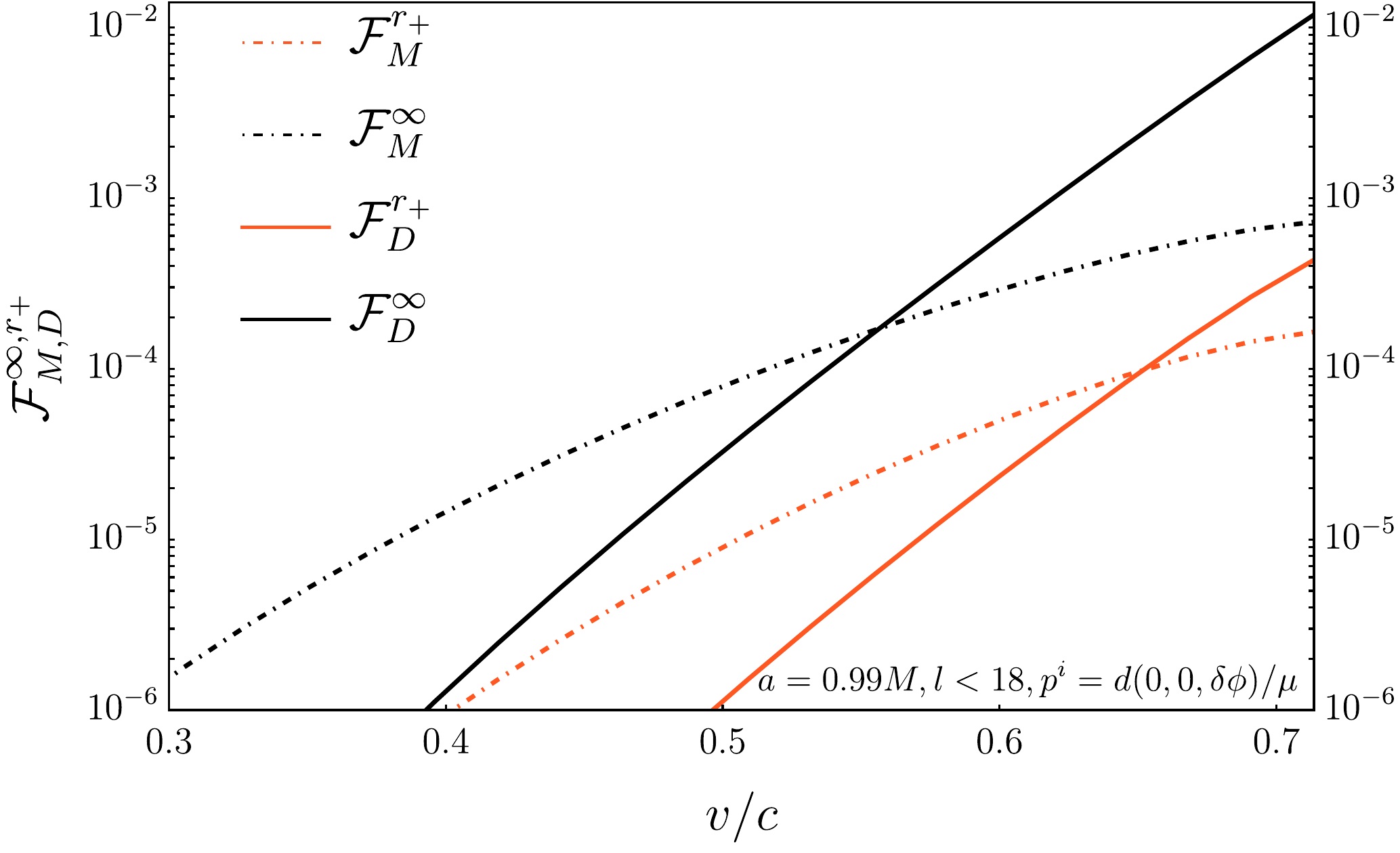}
    \caption{Same as Fig.~\ref{fig:Theta&Mono} but for the case of a dipole moment with $\delta r = \delta \theta = 0$.}
    \label{fig:Phi&Mono}
\end{figure}

\subsection{Estimates for the fundamental dipole in various models}
In order to give a rough estimate of the actual effects of fundamental dipole moments, we need to consider the physical fluxes in Eq.~\eqref{NormFluxes} and plug in realistic values for the (dimensionless) scalar charge and dipole moment.

Using Eq.~\eqref{NormFluxes} and the numerical results of the previous section, we now wish to roughly estimate the minimum value of $p$ that could possibly give detectable effects. An order-of-magnitude estimate can be obtained by computing the ratio between the physical dipolar and monopolar fluxes, $F_D/ \ F_M$, and extract the minimum value of $p$ for which the fluxes are comparable, $F_D/ \ F_M \approx {\cal O}(1)$. 

In the monopolar case, the smallest charge that would lead to a detectable effect for the scalar emission from a monopole is $d \approx 10^{-2}$. This was shown in~\cite{Maselli:2021men} both by computing the GW dephasing due to the scalar emission and by performing a more rigorous parameter estimation.
We can therefore estimate whether the dipole emission is comparable to the flux generated by the monopole setup with this minimum detectable value for the scalar charge, using realistic values of $p$. We also assume a mass-ratio $q=10^{-4}$ in order to minimize the suppression factor between dipole and monopole while remaining well within the extreme mass-ratio limit.

The fluxes ratio reads 
        \begin{equation}
            \frac{F_D}{F_M} = \frac{\mathcal{F}_D}{\mathcal{F}_M} \left ( \frac{ p q}{d}  \right)^2 = 10^{-4} \left(\frac{q}{10^{-4}}\right)^2 \left(\frac{0.01}{d}\right)^2\frac{\mathcal{F}_D}{\mathcal{F}_M}  p^2. 
        \end{equation}
Near the ISCO, for nearly-extremal BHs, the maximum normalized ratio is $ \mathcal{F}_D / \mathcal{F}_M \sim 10^2$, as shown in Fig.~\ref{fig:Radial&Mono}. This leads to
\begin{equation}
     \label{eq:fluxesestimate}        \frac{F_D}{F_M} \sim 10^{-2} \left(\frac{q}{10^{-4}}\right)^2 \left(\frac{0.01}{d}\right)^2  p^2 \,.
\end{equation}
Therefore, in order to require a dipolar flux comparable to the monopolar one, $ p \approx 10$.

To assess whether this value for our fundamental dipole is realistic, we consider a few significant examples.
The first one is the magnetic dipole moment of a neutron star. Of course in this case the dipole moment is due to the electromagnetic field, but we will use the intuition from our scalar dipole as a proxy.
The dipole moment of a neutron star can be estimated as $B R^3$, where $B$ is the typical magnetic field and $R$ the radius of the star~\cite{10.1093/pasj/pst014}. If we assume standard parameters for a magnetar~\cite{Kaspi:2017fwg}, $B \approx 10^{15} G$ and $R \approx 12\mathrel{\textrm{km}}$, we obtain a dipole moment $ p\approx 10^{-2}$.
Thus, from Eq.~\eqref{eq:fluxesestimate}, even in the case of extreme magnetic fields and in the most optimistic scenario the dipole flux is $10^6$ times smaller than the minimum detectable monopole flux.

The same occurs for the dipole moment of the recently constructed globally neutral topological solitons~\cite{Bah:2022yji}. Such solutions are constructed by two opposite charges held at a given distance. While their dipole moment depends on the parameter space, in our units these solutions are characterized by $p \ll 1$~\cite{Bah:2022yji}. 

Furthermore, in the context of quadratic gravity theories with scalar fields coupled to quadratic curvature terms (which most notably include scalar Gauss-Bonnet and dynamical Chern-Simons theories)
dipole hair can grow dynamically~\cite{R:2022tqa}. In this scenario the dipole moment is completely determined by the value of the monopole hair yielding $p \sim d$. In Chern-Simons gravity, dipole hair is proportional to the BH spin and is therefore bounded also in this case~\cite{R:2022tqa}.
Likewise, fundamental dipoles can be produced in theories with spin-induced scalarization~\cite{Dima:2020yac,Berti:2020kgk,Herdeiro:2020wei,Elley:2022ept} at the level of $p \sim d$.

Overall, for generic values of $d$, Eq.~\eqref{eq:fluxesestimate} implies a very large magnitude, $p\approx 10^3 d$, for the dipole flux to be comparable to the monopole one for $q=10^{-4}$. Furthermore, the estimate~\eqref{eq:fluxesestimate} is already very optimistic, as it assumes the smallest detectable value of a scalar charge $d$, a moderate mass-ratio $q=10^{-4}$, and the maximum possible normalized ratio $\mathcal{F}_D / \mathcal{F}_M$, obtained near the ISCO of a nearly extremal central BH. If these assumptions are relaxed (i.e., for smaller mass ratios, larger values of the charge, moderately spinning BHs, and less relativistic orbits), the flux ratio is even more severely suppressed.

We conclude that, for typical values of the dipole moment predicted in various models, the effect of a dipole flux is negligible.

 \subsection{Mixed case: charge+dipole} \label{sec:mixed}
Until now, we have neglected the presence of a putative scalar charge and focused purely on the dipole contribution, assuming a globally neutral secondary. One might wonder if, in a scenario where both a scalar charge and a dipole are included, the corrections to the total flux due to the dipole are more significant. Indeed, as already mentioned, in this scenario the lowest order contribution from the dipole scales as $q^3$ instead of $q^4$. In this section, we will therefore compute the fluxes in this mixed case and show that, also in this case, one needs large values of the dipole moment ($p \approx 1$) in order to have appreciable deviations from the purely monopolar case.

\begin{figure*}[h]
    \centering
    \includegraphics[width = 0.95\textwidth]{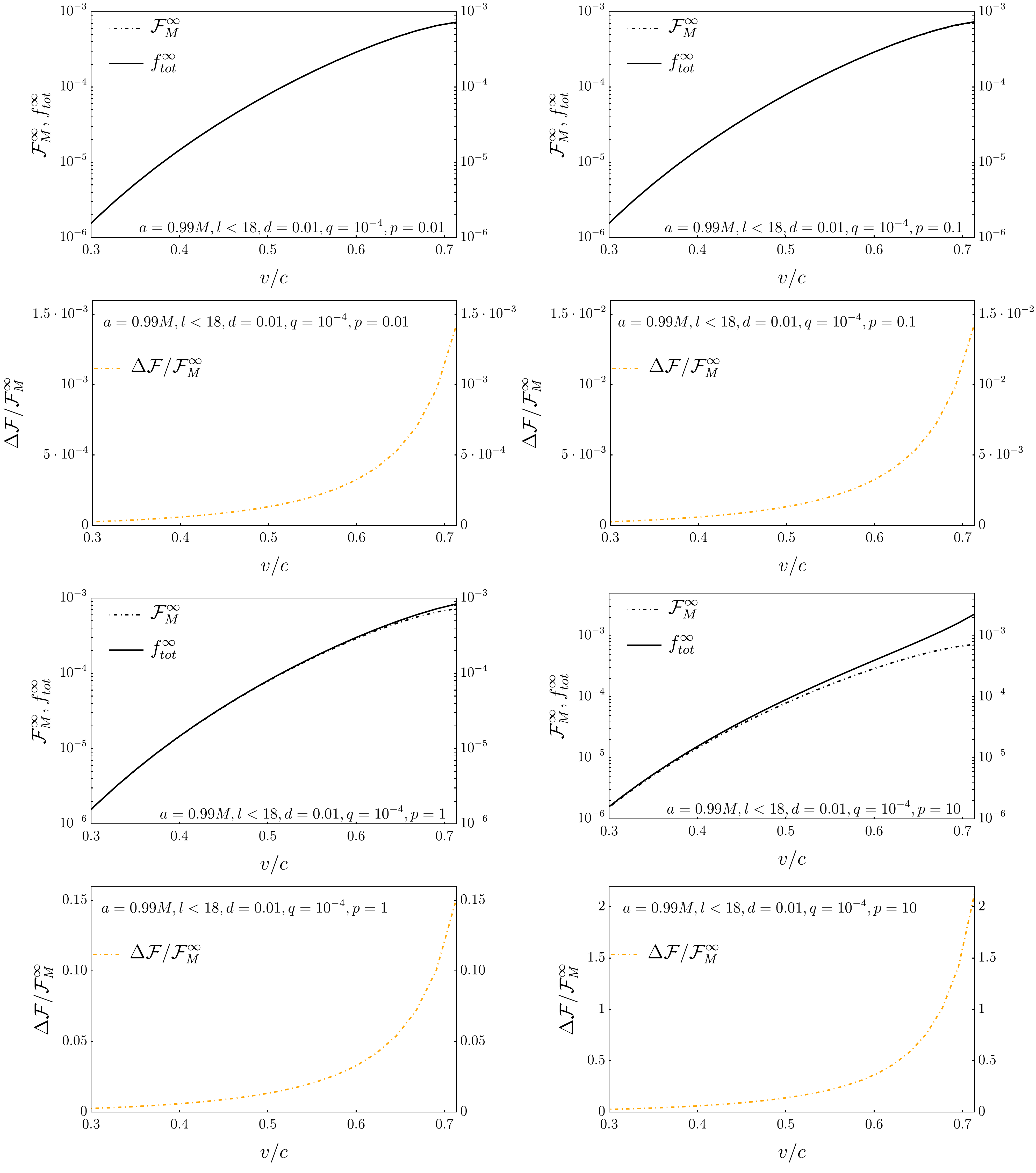}
    \caption{First and third row of panels: mixed case fluxes normalized by $(4 \pi)^2 \mu^2 d^2$ (solid lines) compared to normalized monopolar fluxes (dashed lines) for increasing values of $p$ ($p=0.01,0.1,1,10$).  %
    Second and fourth row of panels: corresponding relative difference between the fluxes at infinity for the same values of $p$. The spin of the BH is fixed at $a = 0.99 M$ and the orientation of the dipole is given by $p^{i} = d (\delta r,0,0) / \mu$.}
    \label{fig:SmallRelativeDifference}
\end{figure*}

We have studied the relative difference in the same optimistic scenario of the previous section, i.e. $a = 0.99 M$, $q = 10^{-4}$ , $d = 10^{-2}$, by considering different values of the dipole moment, namely $p=10^{-2},0.1, 1, 10$. Moreover, we have considered the purely radial dipole case $p^i = d (\delta r, 0, 0) / \mu$, since our previous analysis showed that in this orientation the dipole contribution is maximized.
Figure~\ref{fig:SmallRelativeDifference} shows the total fluxes (continuous lines) and the purely monopolar ones (dashed lines) as functions of the orbital velocity and for increasing values of $p$. Below each panel we also show the corresponding relative difference between monopole+dipole and purely monopole cases.

It is clear that also in this mixed case the presence of a nonvanishing dipole moment has a small impact on the total emission. For realistic values as $p\approx0.01$, the total flux is larger than the the purely monopolar case at most by $\approx 0.1\%$. If we assume a larger dipole moment, $p\approx1$, the maximum deviation from the purely monopolar case is at most $15\%$, whereas for $p\approx10$ the monopolar and dipolar contributions are of the same order. Nevertheless, as already mentioned, such values of $p$ are unrealistic and these corrections are obtained in the most optimistic scenario.

\section{Conclusions}
Motivated by various scenarios predicting globally neutral compact objects endowed with a dipole moment, we have developed a model-agnostic framework to compute the GW emission from a fundamental scalar dipole in EMRIs.

We found that the extra flux associated with the dipole moment is suppressed by the square of the mass ratio relative to the case of fundamental charges, making its detection with EMRIs very challenging for the typical values of the dipole predicted in various models.
Even in the most optimistic scenarios, we estimated that, as long as the dimensionless dipole moment $p\lesssim10$, its effect would be negligible for LISA.

This negative conclusion is based on a simple comparison between the dipole and monopole fluxes. Although the strong suppression suggests that our conclusion is solid, it would be important to confirm this expectation through a proper parameter estimation, along the lines of~\cite{Maselli:2021men,Zhang:2023vok} for the case of fundamental charges.
In our case, however, this would come with the extra cost of properly taking into account the dipole precession during the evolution. Indeed, we have found that the only case in which precession is absent (when the dipole is parallel to the spin of the primary) is also the one in which dipole emission is more suppressed.
For the most promising cases (any other orientation of the dipole) one needs to consistently evolve the dynamics of the dipole moment, similarly to the case of a secondary spin~\cite{Dolan:2013roa,Lukes-Gerakopoulos:2017vkj,Akcay:2016dku,Akcay:2017azq}.
It is also possible that precession helps make the effects of the dipole moment more prominent, as recently found in the context of post-Newtonian theory for comparable-mass binaries~\cite{Loutrel:2023boq}.

Although in our settings the effect of a fundamental dipole on the EMRI fluxes seems pessimistically small, for the same reason we estimate that if the secondary is endowed with both a charge and a dipole, the effect of the latter are typically negligible 
for constraints on and detectability of the former. This suggests that the estimates in~\cite{Maselli:2020zgv,Maselli:2021men,Barsanti:2022ana,Barsanti:2022vvl,Liang:2022gdk,Zhang:2023vok} should be robust against the inclusion of extra dipole effects.
The mixed case is less suppressed by the mass ratio and is in fact very similar to that of an ordinary EMRIs with a spinning secondary~\cite{Piovano:2020zin,Piovano:2022fsy} for which the secondary spin is indeed not measurable with LISA, at least when neglecting precession~\cite{Piovano:2021iwv}.
In this context it would be interesting to include our effect in a more accurate self-force model, see~\cite{Spiers:2023cva} for very recent related work.

Finally, while the case of a fundamental scalar dipole might be interesting on its own in the context of modified gravity and physics beyond the Standard Model, we have also used it as a proxy for an intrinsic electromagnetic dipole, which is of direct astrophysical interest for magnetars. It would be very interesting to extend our work to the vector case, generalizing~\cite{Liang:2022gdk,Zhang:2023vok} to the case of fundamental vector dipoles.

\begin{acknowledgments}
We are grateful to Nick Loutrel for interesting discussion.
We acknowledge financial support provided under the European
Union's H2020 ERC, Starting Grant agreement no.~DarkGRA--757480 and support under the MIUR PRIN (Grant 2020KR4KN2 “String Theory as a bridge between Gauge Theories and Quantum Gravity”) and FARE (GW-NEXT, CUP: B84I20000100001, 2020KR4KN2) programmes.
We acknowledge additional financial support provided by "Progetti per Avvio alla Ricerca - Tipo 1", protocol number AR1221816BB60BDE.
\end{acknowledgments}

\bibliography{refs}
\end{document}